\documentclass[cameraready]{Interspeech}

\title{Rethinking Speaker Embeddings for Speech Generation: \\ Sub-Center Modeling for Capturing Intra-Speaker Diversity} 

\author[affiliation={1}, orcid=0000-0003-4593-9057]{Ismail Rasim}{Ulgen}
\author[affiliation={2}, orcid=0000-0003-1382-9929]{John H. L.}{Hansen}
\author[affiliation={3}, orcid=0000-0002-4075-4072]{Carlos}{Busso}
\author[affiliation={1},orcid=0000-0001-8078-3305]{Berrak}{Sisman} 

\address{
    $^1$ Center for Language and Speech Processing (CLSP), Johns Hopkins University, USA \\
    $^2$ Center for Robust Speech Systems, The University of Texas at Dallas, USA \\
    $^3$ Language Technologies Institute, Carnegie Mellon University, USA 
}

\email{iulgen1@jhu.edu, john.hansen@utdallas.edu, busso@cmu.edu, sisman@jhu.edu}

\usepackage{comment}
\usepackage{amsmath,graphicx}
\usepackage{adjustbox}
\usepackage{cite}
\usepackage{booktabs}
\usepackage{subcaption}

\newcommand\blfootnote[1]{%
  \begingroup
  \renewcommand\thefootnote{}\footnote{#1}%
  \addtocounter{footnote}{-1}%
  \endgroup
}

\begin{document}

\maketitle
\blfootnote{\textbf{Speech samples:} \url{https://choughtotem.github.io/subcentervc_demo/}}
\begin{abstract}

Modeling speech variation is key to natural, expressive generation. Speaker embeddings are commonly used to condition personalized speech systems, but they are typically trained for speaker recognition, where intra-speaker variability is suppressed and inter-speaker separation is maximized. This objective leads to overly compact representations that may discard variations crucial for generation. We revisit this design choice and propose a sub-center modeling framework for speaker embeddings. Instead of a single prototype per speaker, we learn multiple sub-centers during discriminative training, allowing utterances to align with different prototypes. This strategy preserves structured intra-speaker variability while maintaining discriminability. In zero-shot voice conversion, our method improves intelligibility, increases pitch variability, achieves higher naturalness ratings, and retains strong speaker verification performance.
\end{abstract}

\section{Introduction}
Speaker embeddings, originally developed for recognition tasks \cite{x-vector, ecapa, ge2e}, are typically learned via large-scale speaker classification objectives that encourage small intra-speaker variation and large inter-speaker separation. Due to their strong generalization, these embeddings are now widely used as conditioning signals in downstream speech generation systems such as text-to-speech (TTS) \cite{yourtts,yamagishitts} and voice conversion (VC) \cite{autovc,wavthruvec}.

Zero-shot, multi-speaker generation methods \cite{autovc, yourtts, wavthruvec, yamagishitts} can synthesize speech for unseen speakers from a reference utterance, enabling flexible personalization. In TTS, text is mapped to speech; in VC, the source speech is transformed into the target speaker’s voice while preserving linguistic content \cite{berisha, free-vc, polyak21_interspeech, popov2022diffusionbased}. In both settings, speaker embeddings are used to convey speaker identity. Beyond identity, however, natural-sounding synthesis requires modeling expressive intra-speaker variation such as prosody, speaking style, and emotion \cite{sismanoverview, yamagishi_overview, du2025naturalvoices}. Recently in voice conversion, we see advances in expressive and emotional speech generation \cite{zhou2021seen, zhou2022emotional}, in which models use speaker embeddings together with expressive cues while modeling the generation. These approaches highlight the importance of modeling intra-speaker diversity during generation.

Notably, there is a fundamental mismatch between recognition and generation. Speaker recognition seeks to distinguish a given speaker from others by minimizing intra-class variance and maximizing inter-class separation \cite{intra-class_min,metric-sv}. However, reducing intra-class variance often suppresses meaningful variability across utterances of the same speaker, resulting in embeddings that lack expressiveness. Conventional classification-based embedding models represent each speaker with a single prototype in the embedding space, encouraging all utterances to collapse toward one center \cite{softtriple}. While this structure benefits discrimination, it is misaligned with the objective of generation, where intra-speaker variability (e.g., prosody, style, emotion) is part of the desired output distribution. Embeddings that are highly effective for recognition may be suboptimal for generation tasks.

Existing advances in speaker embedding learning continue to prioritize minimizing intra-class variance to improve recognition performance \cite{generalized}, and embeddings used in generation systems are commonly adopted from this recognition-driven paradigm with minimal reconsideration. In contrast, we treat intra-speaker variance as a \emph{generation-relevant} design dimension and ask: \textit{Can relaxing overly compact speaker representations yield embeddings that better support expressive speech generation, without sacrificing speaker discriminability?}

This work addresses the recognition–generation mismatch by revisiting the structural assumption in conventional speaker embedding training. Rather than representing each speaker with a single prototype, we introduce a sub-center formulation that learns multiple prototypes per speaker within a discriminative objective. This strategy allows different utterances to align with different sub-centers, preserving meaningful intra-speaker variation in embeddings while maintaining speaker discriminability. We evaluate the proposed embeddings in a zero-shot, multi-speaker voice conversion setting as a representative speech generation task. Our results demonstrate that increasing intra-speaker variability in the embedding space leads to more natural and prosodically expressive synthesized speech, without degrading speaker recognition performance. These findings suggest that embedding compactness, traditionally optimized for recognition, should be reconsidered when embeddings are used in generation tasks. 

Our contributions are as follows: 
\begin{itemize}
    \item We introduce a generation-oriented view of speaker embedding learning that reframes intra-speaker variance as a design parameter rather than a quantity to suppress.
    \item We propose a sub-center training formulation that preserves structured intra-speaker variability within a discriminative objective.
    \item Through VC, we demonstrate that richer intra-speaker variability improves naturalness and prosodic expressiveness while maintaining strong speaker verification performance.
\end{itemize}

\begin{figure*}[t]
\hspace*{-8mm}
    \centering
    \scalebox{0.25}{\includegraphics{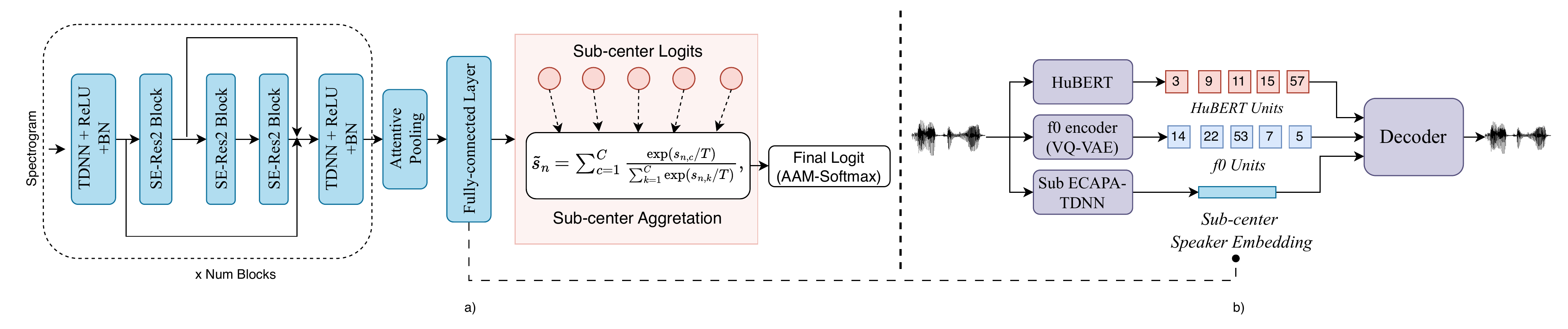}}
    
    \caption{a) Proposed sub-center modeling applied to the traditional ECAPA-TDNN network. b) VC framework using the proposed embeddings}
    \label{fig:diagram}
    
\end{figure*}
\section{Related Work}
Sub-center classification has been explored in computer vision. For example, Qian et al. \cite{softtriple} and Zhang \& Gong \cite{fixed_subcenter} improved fine-grained retrieval by representing each class with multiple sub-centers. Müller et al. \cite{subclass_distill} showed sub-centers help teacher models discover latent subclasses, while Deng et al. \cite{sub-center_arcface} used them in face recognition to separate noisy from clean samples. In all these cases, the objective was a discriminative task: sub-centers were a means to refine classification boundaries. In speaker recognition, sub-centers have been applied mainly for verification under noisy labeled or unlabeled conditions \cite{idlab, speakin}. The goal was to improve the classification by singling-out noisy labeled examples. Importantly, the embeddings were still optimized to minimize intra-speaker variance, as variation is treated as a source of error for recognition \cite{positive_samp,max_sep}. To our knowledge, no prior work has explored the perspective of intra-class variance of embeddings in generation tasks such as VC \cite{sismanoverview,oytun_overview} and utilized sub-center embeddings explicitly for this purpose. Our work is the first to repurpose sub-center modeling to preserve intra-speaker variation for speech generation, rather than suppress it for recognition. We provide both theoretical motivation and empirical evidence that embeddings with higher intra-class variance produce more natural, expressive speech while retaining speaker identity.

\section{Sub-center Modeling for Speaker Embeddings}
Conventional speaker embeddings are designed with recognition in mind, where the objective is to minimize intra-class variability and maximize inter-class separation. While effective for discrimination, this design limits the natural diversity within a speaker’s voice, such as prosody, emotion, and style. For downstream generation tasks, however, such intra-speaker variation is not noise but a critical factor for producing natural and expressive speech. We, therefore, propose a sub-center modeling strategy for speaker embeddings, in which each speaker is represented by multiple prototypes rather than a single class center. This formulation allows embeddings to capture diverse realizations of the same speaker while retaining discriminability. Our framework is broadly applicable to speech generation tasks, and we demonstrate its effectiveness through voice conversion as a representative case study. 

\subsection{Speaker Embeddings with Single Class-center}
We adopt the Emphasized Channel Attention, Propagation, and Aggregation in Time-Delay Neural Network (ECAPA-TDNN) \cite{ecapa} as a representative speaker embedding network, due to its strong performance and widespread use in recognition tasks. ECAPA-TDNN encodes a speech utterance into a fixed-dimensional embedding vector, which is then classified using an additive angular margin softmax (AAM-Softmax) objective with speaker labels. The classifier head is parameterized by a weight matrix $W \in \mathbb{R}^{L \times N}$, where $L$ is the embedding dimension and $N$ is the number of training speakers. Each column $w_n \in \mathbb{R}^{L}$ can be interpreted as the prototype, or class center, for speaker $n$. The AAM-Softmax loss for an embedding $x_i$ of speaker $y$ is
\setlength{\abovedisplayskip}{3pt}
\setlength{\belowdisplayskip}{3pt}
\begin{equation}
\mathcal{L} = - \log \frac{e^{k \cos(\theta_{y} + m)}}{e^{k \cos(\theta_{y} + m)} + \sum\limits_{j=1,  j \neq y}^{N} e^{k \cos(\theta_{j})}},
\end{equation}
where $\cos(\theta_{j}) = w_j^\top x_i$, $m$ is the angular margin, and $k$ is a scale factor. The objective encourages embeddings of the same speaker to cluster tightly around a single center $w_y$, while remaining separable from other speakers. This single-center formulation is well suited for recognition, where the goal is invariance to intra-speaker variability. However, collapsing all utterances of a speaker to one point in the embedding space inevitably discards natural variation due to prosody, style, or emotion. As a result, single-center embeddings are suboptimal for generation tasks, which require these variations to be preserved.

\subsection{The Proposed Sub-center Modeling}
An illustration of the sub-center ECAPA-TDNN is shown in Fig.~\ref{fig:diagram}. To preserve intra-speaker variation while maintaining speaker discriminability, we extend the AAM-Softmax objective in ECAPA-TDNN to include multiple sub-centers per speaker class, and repurpose these sub-centers for more variation. The classifier weights become 
$W_c \in \mathbb{R}^{L \times N \times C}$, where $L$ is the embedding dimension, $N$ is the number of speaker classes, and $C$ is the number of sub-centers per class. For each speaker $n$, we maintain sub-centers $\{w_{n,1}, w_{n,2}, \dots, w_{n,C}\}$. Given an embedding $x_i$ of speaker $y_i$, the similarity to the $c$-th sub-center of class $n$ is $s_{n,c} = w_{n,c}^\top x_i$. We aggregate similarities within each class using a temperature-scaled softmax weights:

\begin{equation}
\alpha_{n,c} = \frac{e^{s_{n,c}/T}}{\sum_{k=1}^{C} e^{s_{n,k}/T}}, 
\end{equation}
\begin{equation}
\tilde{s}_n = \sum_{c=1}^{C} \alpha_{n,c} s_{n,c}.
\end{equation}
where $\alpha_{n,c}$ is similarity weight for sub-center $c$ of class $n$, and $\tilde{s}_n$ is the aggregated similarity for class $n$. 

The sub-center AAM-Softmax loss is then

\begin{equation}
\mathcal{L}
= -\log\frac{e^{\,k\cos(\tilde{\theta}_{y_i}+m)}}
{\,e^{\,k\cos(\tilde{\theta}_{y_i}+m)} + \displaystyle\sum\limits_{j\neq y_i} e^{\,k\cos(\tilde{\theta}_j)}\,}.
\end{equation}
where $\cos(\tilde{\theta}_n) = \tilde{s}_n$ is the aggregated sub-class similarities for class $n$, $m$ is the angular margin, and $k$ is a scaling factor.

This proposed formulation allows each utterance to selectively align with one or more sub-centers rather than being forced toward a single prototype. As a result, embeddings can capture intra-speaker variations such as prosody and style, while still being trained under the same discriminative classification objective.




\subsection{Voice Conversion}

We assess the proposed embeddings in a generation setting using voice conversion (VC) \cite{sismanoverview} as a downstream task. VC modifies the speaker identity of an utterance while preserving its linguistic content. We adopt the speech-resynthesis framework of Polyak et al. \cite{polyak21_interspeech}, which factorizes speech into linguistic units, pitch, and speaker identity, and reconstructs the waveform with a HiFi-GAN vocoder \cite{hifigan} (Fig.~\ref{fig:diagram}). At inference, linguistic and pitch features come from a source utterance, and the target identity is provided by a reference embedding extracted from a reference utterance.

For speaker representation, we replace the baseline ECAPA-TDNN embeddings with our sub-center embeddings (Section~3.2). The embedding $s \in \mathbb{R}^{192}$, extracted from a reference utterance, must capture not only timbre but also prosody and style for natural speech generation. Linguistic content is modeled with discrete HuBERT units \cite{polyak21_interspeech, hubert} obtained by $k$-means clustering, and pitch is represented by discrete units derived from normalized $F_0$ contours using the VQ-VAE \cite{dhariwal2020jukebox} strategy. The modified HiFi-GAN decoder then generates expressive speech conditioned on linguistic units, pitch units, and the proposed speaker embedding.

\section{Experimental Setup}
\subsection{Datasets}
The proposed speaker embeddings are trained on VoxCeleb2 \cite{voxceleb2}.  For the VC experiments, we use the VCTK corpus \cite{vctk}, which consists of 110 English speakers, each with approximately 400 utterances. We randomly select 90 speakers for training, while the remaining 20 speakers are utilized for zero-shot VC experiments as unseen speakers.

\vspace{-2mm}
\subsection{Training \& Implementation}
The baseline ECAPA-TDNN is trained using the SpeechBrain recipe \cite{speechbrain}, which we extend to support sub-center modeling by modifying the ECAPA-TDNN architecture. We use the Adam optimizer with a base learning rate of $1\text{e}{-4}$ and a cyclic schedule. The batch size is 32, and we use online augmentation (noise, reverberation) following Desplanques et al. \cite{ecapa}. AAM-Softmax parameters are set to margin $m = 0.2$ and scale $s = 30$. For VC training, we adopt the speech-resynthesis framework \cite{polyak21_interspeech}, modified to use ECAPA-TDNN embeddings. Linguistic features are obtained from the 6th layer of HuBERT, clustered via k-means ($K=100$) trained on LibriSpeech-clean-100 \cite{librispeech}. Pitch is extracted using the Dio algorithm \cite{dio} with default parameters. All encoders (linguistic, f0, speaker) are pre-trained and frozen during HiFi-GAN vocoder training.


\subsection{Evaluation}

\subsubsection{Speaker Verification \& Intra-class Variance}

We evaluate the intra-class variance of standard ECAPA-TDNN and our proposed sub-center embeddings. To compare across embedding spaces, we use the ratio of intra-to inter-class variance as a nominal measure. We calculate intra-class variance as
\begin{equation}
    \sigma_{intra-class}^2 = \frac{\sum^N (f(x_{s,i},\Tilde{x_{s}}) - \mu_{intra})^2}{N}
\end{equation}
where $x_{s,i}$ is the $i^th$ speaker embedding from speaker $s$, $\Tilde{x_s}$ is the mean of all embeddings from speaker $s$, and $f$ is the cosine distance function. $\mu_{intra}$ is the mean of all intra-class cosine distances, and $N$ is the total number of examples. We define the inter-class variance as:
\begin{equation}
\sigma_{\text{inter-class}}^2 = \frac{\sum^{N\times(S-1)} (f(x_{s,i},\Tilde{x_{s'}}) - \mu_{\text{inter}})^2}{N\times(S-1)}
\end{equation}
where we measure the distance between the $i$-th speaker embedding from speaker $s$ and every other speaker's mean embedding $s'$ different from $s$. As the final inter-class variation measure, we report the ratio $\sigma_{\text{intra-class}}^2 / \sigma_{\text{inter-class}}^2$ throughout the paper. For speaker verification, we generated 20M trials from 110 VCTK speakers and use voxceleb1-e test set\cite{voxceleb2}. We measured equal-error-rate (EER) using cosine similarity between pairs.
\begin{table*}[t]
\centering
\caption{Speaker recognition and intra-/inter-class variance.}
\label{tab:sv}
\vspace{-0.3cm}
\begin{tabular}{lcccc}
\toprule
& \multicolumn{2}{c}{VCTK} & \multicolumn{2}{c}{VoxCeleb1-E} \\
\cmidrule(lr){2-3} \cmidrule(lr){4-5}
Embedding & EER (\%) $\downarrow$ & Intra/inter-class var $\uparrow$ & EER (\%) $\downarrow$ & Intra/inter-class var $\uparrow$ \\
\midrule
ECAPA-TDNN~\cite{ecapa}                                 & 1.71 & 0.42 & 1.46 & 0.66 \\
Sub-center ECAPA-TDNN ($C=10, T=0.1$)           & \textbf{1.47} & 0.36 & 1.33 & 0.65 \\
Sub-center ECAPA-TDNN ($C=10$)                  & 1.50 & 0.45 & \textbf{1.15} & 0.82 \\

Sub-center ECAPA-TDNN ($C=20$)                  & 1.55 & \textbf{0.47} & 1.21 & \textbf{0.91} \\
\bottomrule
\end{tabular}
\vspace{-0.3cm}
\end{table*}

\begin{table}[t]
\centering
\caption{Analysis of variation in synthesized speech.}
\label{tab:variation}
\vspace{-0.3cm}
\fontsize{8}{11}\selectfont
\begin{tabular}{l@{\hspace{0.07cm}}c@{\hspace{0.07cm}}|@{\hspace{0.07cm}}c@{\hspace{0.07cm}}|@{\hspace{0.07cm}}c}
\toprule
Method & $F0$ std $\uparrow$ & $F0$ range $\uparrow$ & Var $\uparrow$ \\
\midrule
VC w/ ECAPA-TDNN~\cite{ecapa} & 8.03 & 52.37 & 0.147 \\
VC w/ Sub ECAPA-TDNN ($C$=20) & \textbf{10.25} & \textbf{57.09} & \textbf{0.167} \\
\bottomrule
\end{tabular}
\end{table}

\begin{table}[t]
\centering
\caption{Objective evaluations for VC.}
\label{tab:vc_obj}
\vspace{-0.3cm}
\fontsize{8}{11.5}\selectfont
\scalebox{0.95}{
\begin{tabular}
{l@{\hspace{0.1cm}}c@{\hspace{0.1cm}}c@{\hspace{0.1cm}}c@{}}
\toprule
Method & WER $\downarrow$ & CER $\downarrow$ & SECS (\%) $\uparrow$ \\
\midrule
VC w/ ECAPA-TDNN~\cite{ecapa} & 14.84 & 6.82 & 64.04 \\
VC w/ Sub ECAPA-TDNN ($C$=10) & 14.65 & 6.72 & 64.14 \\
VC w/ Sub ECAPA-TDNN ($C$=10, $T$=0.1) & 14.32 & 6.67 & \textbf{65.86} \\
VC w/ Sub ECAPA-TDNN ($C$=20) & \textbf{13.93} & \textbf{6.41} & 64.59 \\
\bottomrule
\end{tabular}}
\vspace{-3mm}
\end{table}

\begin{table}[t]
\centering
\caption{Subjective evaluation results for VC (95\% CI).}
\label{tab:subj}
\fontsize{7.6}{11}\selectfont
\vspace{-0.3cm}
\hspace{-1mm}
\begin{tabular}{@{}l@{\hspace{0.1cm}}c@{\hspace{0.1cm}}c@{}}
\toprule
Method & MOS $\uparrow$ & SMOS $\uparrow$ \\
\midrule
Ground Truth & $4.65 \pm 0.09$ & -- \\
\midrule
VC w/ ECAPA-TDNN~\cite{ecapa} & $2.94 \pm 0.12$ & $2.65 \pm 0.13$ \\
VC w/ Sub ECAPA-TDNN ($C$=10, $T$=0.1) & $2.89 \pm 0.13$ & $2.76 \pm 0.13$ \\
VC w/ Sub ECAPA-TDNN ($C$=20) & $\mathbf{3.18 \pm 0.12}$ & $\mathbf{2.88 \pm 0.13}$ \\
\bottomrule
\end{tabular}
\end{table}

\subsubsection{VC Evaluation}

\footnotetext[1]{https://huggingface.co/facebook/wav2vec2-large-960h-lv60-self}
\footnotetext[2]{https://github.com/resemble-ai/Resemblyzer}
We evaluate our approach for VC task using both objective and subjective metrics. In the VC setting, we generate ~20,000 converted utterances by randomly selecting a source utterance and a reference utterance from unseen speakers in each VC trial, covering a wide range of different source and reference utterance combinations. This extensive sampling allows intra-speaker variation in the embedding space to manifest across diverse conversion scenarios. Objective evaluation includes word error rate (WER) and character error rate (CER) \cite{diagnostics} from a state-of-the-art (SOTA) automatic speech recognition (ASR) model\footnotemark[1] \cite{wav2vec2}, and speaker embedding cosine similarity (SECS) using a pre-trained d-vector model\footnotemark[2] \cite{ge2e}, across converted utterances. For subjective evaluations, we conduct MOS \cite{yamagishi_overview} for naturalness, SMOS \cite{yourtts} for speaker similarity, and ABX tests \cite{sismanoverview} for prosody (intonation, stress, rhythm) naturalness, using 120 samples rated by 12 participants.

\section{Results}

\subsection{Speaker Verification \& Intra-class Variance }
We evaluated our sub-center speaker embeddings using different numbers of sub-centers per class, experimenting with $C = 10$ and $C = 20$, following the setup in Qian et al. \cite{softtriple}. In addition, we tested two temperature values for sub-center logit aggregation: no temperature scaling ($T=1$) and a small temperature ($T=0.1$). Table~\ref{tab:sv} reports the EER for speaker verification as well as the intra-/inter-class variance ratio ($var$), which we use as a normalized measure of intra-speaker variability. The results show that sub-center modeling with $T=1$ achieves higher intra-class variance compared to the standard ECAPA-TDNN, indicating richer embedding representations. Importantly, despite the increased variance, the sub-center models also yield improved EERs, demonstrating that discriminative power is not compromised. These findings suggest that sub-center modeling enables more effective modeling of complex intra-speaker distributions while preserving, or even improving, speaker verification performance.  

Interestingly, sub-center modeling with a low temperature ($T=0.1$) yields lower intra-class variance than the baseline ECAPA-TDNN. A small temperature makes sub-center selection overly confident, causing the model to rely on only a few centers—a behavior also noted in prior work \cite{fixed_subcenter,subclass_distill}. This result underscores the role of temperature in controlling sub-center utilization. Notably, the $T=0.1$ configuration also achieves the best verification performance in VCTK and second-best in voxceleb1-e, suggesting that tighter clustering still benefits recognition, although reduced variability may limit its usefulness for generation tasks.

\subsection{Variation Analysis in Voice Conversion}
\vspace{-2mm}
To examine how the intra-class variance of embeddings affects the generated speech, we analyze variation in the converted utterances using utterance-level $F0$ standard deviation, $F0$ range, and the intra/inter-class variance ratio of d-vector embeddings, denoted by Var, extracted from the synthesized speech. The $F0$ statistics reflect pitch diversity, while embedding variance measures the spread of speaker representations after conversion. As shown in Table~\ref{tab:variation}, Sub-center ECAPA-TDNN ($C=20$) produces higher pitch variability and embedding variance than the baseline. These results suggest that relaxing intra-class compactness during training leads to more diverse synthesized outputs.

\subsection{Objective Evaluation of Voice Conversion}
Table~\ref{tab:vc_obj} shows that the sub-center ECAPA-TDNN with $C=20$ (highest intra-class variance) achieves the lowest WER and CER, indicating better intelligibility and synthesis quality, and also improves SECS over the baseline. In contrast, the lowest-variance model ($C=10, T=0.1$) yields the highest SECS, reflecting stronger identity matching. This result reveals a trade-off: higher variance favors intelligibility, while lower variance favors speaker similarity. Notably, both sub-center configurations outperform the baseline across all metrics.

\subsection{Subjective Evaluation of Voice Conversion}
\begin{figure}[t]
    \centering
    \includegraphics[width=\linewidth]{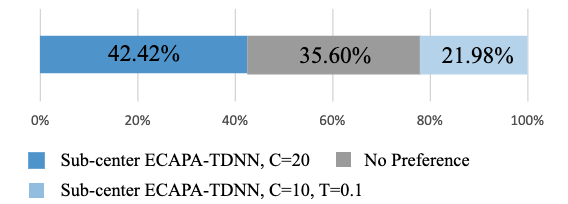}
    \caption{ABX prosody preference results between the VC with embeddings having highest and lowest intra-class variance.}
    \label{fig:proso}
    \vspace{-0.4cm}
\end{figure}

Subjective results in Table~\ref{tab:subj} and Fig.~\ref{fig:proso} align with the objective findings. Embeddings with the highest intra-class variance achieve the best MOS for naturalness, with statistical significance confirmed by a one-tailed paired t-test ($p<0.05$). This configuration also gives the highest similarity MOS and outperforms lower-variance models in the ABX prosody test. Overall, sub-center modeling improves over the baseline in nearly all evaluations: while lower-variance embeddings aid speaker discrimination, higher-variance embeddings are clearly better at capturing prosodic variation and producing more natural, expressive speech.

\vspace{-3mm}
\section{Conclusions}
Speaker embeddings have traditionally been optimized for recognition, where minimizing intra-speaker variability is desirable for discrimination. However, these same embeddings are increasingly used as conditioning inputs in speech generation systems, where variability contributes to naturalness and expressiveness. In this work, we revisited this recognition-driven assumption and introduced a sub-center modeling framework that relaxes overly compact speaker representations while preserving identity discrimination. Through analysis and voice conversion experiments, we showed that embeddings with richer intra-class variability produce more natural and expressive speech without degrading verification performance. These findings highlight the importance of aligning embedding design with downstream generative objectives and suggest a complementary direction for developing speaker representations tailored to speech generation.

\section{Acknowledgments}
This work was supported by the National Science Foundation (NSF) CAREER Award IIS-2533652. The views and conclusions contained herein are those of the authors and should not be interpreted as necessarily representing the official policies of the U.S. Government. The U.S. Government is authorized to reproduce and distribute reprints for governmental purposes, notwithstanding any copyright annotation therein.
\section{Generative AI Use Disclosure}
Generative AI tools were employed solely for language polishing of text written by the authors. These tools were not used to generate scientific content, results, experimental designs, analyses, or conclusions. All authors are responsible for the full content of this paper and consent to its submission.

\begingroup

\bibliographystyle{IEEEtran}
\bibliography{mybib}

@inproceedings{ecapa,
  author={Brecht Desplanques and Jenthe Thienpondt and Kris Demuynck},
  title={{ECAPA-TDNN: Emphasized Channel Attention, Propagation and Aggregation in TDNN Based Speaker Verification}},
  year=2020,
  booktitle={Proc. Interspeech 2020},
  pages={3830--3834},
  doi={10.21437/Interspeech.2020-2650}
}

@INPROCEEDINGS{x-vector,
  author={Snyder, David and Garcia-Romero, Daniel and Sell, Gregory and Povey, Daniel and Khudanpur, Sanjeev},
  booktitle={2018 IEEE International Conference on Acoustics, Speech and Signal Processing (ICASSP)}, 
  title={X-Vectors: Robust DNN Embeddings for Speaker Recognition}, 
  year={2018},
  volume={},
  number={},
  pages={5329-5333},
  doi={10.1109/ICASSP.2018.8461375}}

@INPROCEEDINGS{ge2e,
  author={Wan, Li and Wang, Quan and Papir, Alan and Moreno, Ignacio Lopez},
  booktitle={2018 IEEE International Conference on Acoustics, Speech and Signal Processing (ICASSP)}, 
  title={Generalized End-to-End Loss for Speaker Verification}, 
  year={2018},
  volume={},
  number={},
  pages={4879-4883},
  doi={10.1109/ICASSP.2018.8462665}}

@article{zhou2022emotional,
title = {Emotional voice conversion: Theory, databases and {ESD}},
journal = {Speech Communication},
volume = {137},
pages = {1-18},
year = {2022},
issn = {0167-6393},
doi = {https://doi.org/10.1016/j.specom.2021.11.006},
url = {https://www.sciencedirect.com/science/article/pii/S0167639321001308},
author = {Kun Zhou and Berrak Sisman and Rui Liu and Haizhou Li}
}

@inproceedings{zhou2021seen,
  title={Seen and unseen emotional style transfer for voice conversion with a new emotional speech dataset},
  author={Zhou, Kun and Sisman, Berrak and Liu, Rui and Li, Haizhou},
  booktitle={ICASSP 2021-2021 IEEE International Conference on Acoustics, Speech and Signal Processing (ICASSP)},
  pages={920--924},
  year={2021},
  organization={IEEE}
}

@article{du2025naturalvoices,
  title={NaturalVoices: A Large-Scale, Spontaneous and Emotional Podcast Dataset for Voice Conversion},
  author={Du, Zongyang and Chandra, Shreeram Suresh and Ulgen, Ismail Rasim and Mahapatra, Aurosweta and Salman, Ali N and Busso, Carlos and Sisman, Berrak},
  journal={arXiv preprint arXiv:2511.00256},
  year={2025}
}

@InProceedings{autovc,
  title = 	 {{A}uto{VC}: Zero-Shot Voice Style Transfer with Only Autoencoder Loss},
  author =       {Qian, Kaizhi and Zhang, Yang and Chang, Shiyu and Yang, Xuesong and Hasegawa-Johnson, Mark},
  booktitle = 	 {Proceedings of the 36th International Conference on Machine Learning},
  pages = 	 {5210--5219},
  year = 	 {2019},
  editor = 	 {Chaudhuri, Kamalika and Salakhutdinov, Ruslan},
  volume = 	 {97},
  series = 	 {Proceedings of Machine Learning Research},
  month = 	 {09--15 Jun},
  publisher =    {PMLR},
  url = 	 {https://proceedings.mlr.press/v97/qian19c.html}
}

@InProceedings{yourtts,
  title = 	 {{Y}our{TTS}: Towards Zero-Shot Multi-Speaker {TTS} and Zero-Shot Voice Conversion for Everyone},
  author =       {Casanova, Edresson and Weber, Julian and Shulby, Christopher D and Junior, Arnaldo Candido and G{\"o}lge, Eren and Ponti, Moacir A},
  booktitle = 	 {Proceedings of the 39th International Conference on Machine Learning},
  pages = 	 {2709--2720},
  year = 	 {2022},
  editor = 	 {Chaudhuri, Kamalika and Jegelka, Stefanie and Song, Le and Szepesvari, Csaba and Niu, Gang and Sabato, Sivan},
  volume = 	 {162},
  series = 	 {Proceedings of Machine Learning Research},
  month = 	 {17--23 Jul},
  publisher =    {PMLR},
  url = 	 {https://proceedings.mlr.press/v162/casanova22a.html}
}

@ARTICLE{sisman,
  author={Sisman, Berrak and Yamagishi, Junichi and King, Simon and Li, Haizhou},
  journal={IEEE/ACM Transactions on Audio, Speech, and Language Processing}, 
  title={An Overview of Voice Conversion and Its Challenges: From Statistical Modeling to Deep Learning}, 
  year={2021},
  volume={29},
  number={},
  pages={132-157},
  doi={10.1109/TASLP.2020.3038524}}

@inproceedings{wavthruvec,
  author={Hubert Siuzdak and Piotr Dura and Pol {van Rijn} and Nori Jacoby},
  title={{WavThruVec: Latent speech representation as intermediate features for neural speech synthesis}},
  year=2022,
  booktitle={Proc. Interspeech 2022},
  pages={833--837},
  doi={10.21437/Interspeech.2022-10797}
}

@article{yamagishitts,
  title={Zero-Shot Multi-Speaker Text-To-Speech with State-Of-The-Art Neural Speaker Embeddings},
  author={Erica Cooper and Cheng-I Lai and Yusuke Yasuda and Fuming Fang and Xin Eric Wang and Nanxin Chen and Junichi Yamagishi},
  journal={ICASSP 2020 - 2020 IEEE International Conference on Acoustics, Speech and Signal Processing (ICASSP)},
  year={2019},
  pages={6184-6188},
  url={https://api.semanticscholar.org/CorpusID:204852286}
}

@article{subclass_distill,
  title={Subclass Distillation},
  author={Rafael M{\"u}ller and Simon Kornblith and Geoffrey E. Hinton},
  journal={ArXiv},
  year={2020},
  volume={abs/2002.03936},
}

@INPROCEEDINGS {softtriple,
author = {Q. Qian and L. Shang and B. Sun and J. Hu and T. Tacoma and H. Li and R. Jin},
booktitle = {2019 IEEE/CVF International Conference on Computer Vision (ICCV)},
title = {SoftTriple Loss: Deep Metric Learning Without Triplet Sampling},
year = {2019},
volume = {},
issn = {},
pages = {6449-6457},
doi = {10.1109/ICCV.2019.00655},
url = {https://doi.ieeecomputersociety.org/10.1109/ICCV.2019.00655},
publisher = {IEEE Computer Society},
address = {Los Alamitos, CA, USA},
month = {nov}
}

@article{fixed_subcenter,
  title={The Fixed Sub-Center: A Better Way to Capture Data Complexity},
  author={Zhemin Zhang and Xun Gong},
  journal={ArXiv},
  year={2022},
  volume={abs/2203.12928},

}

@InProceedings{sub-center_arcface,
author="Deng, Jiankang
and Guo, Jia
and Liu, Tongliang
and Gong, Mingming
and Zafeiriou, Stefanos",
editor="Vedaldi, Andrea
and Bischof, Horst
and Brox, Thomas
and Frahm, Jan-Michael",
title="Sub-center ArcFace: Boosting Face Recognition by Large-Scale Noisy Web Faces",
booktitle="Computer Vision -- ECCV 2020",
year="2020",
publisher="Springer International Publishing",
address="Cham",
pages="741--757",
isbn="978-3-030-58621-8"
}

@article{idlab,
  title={The IDLAB VoxCeleb Speaker Recognition Challenge 2020 System Description},
  author={Jenthe Thienpondt and Brecht Desplanques and Kris Demuynck},
  journal={ArXiv},
  year={2020},
  volume={abs/2010.12468},
  url={https://api.semanticscholar.org/CorpusID:260516427}
}

@inproceedings{speakin,
  title     = {{The SpeakIn Speaker Verification System for Far-Field Speaker Verification Challenge 2022}},
  author    = {Yu Zheng and Jinghan Peng and Yihao Chen and Yajun Zhang and Jialong Wang and Min Liu and Minqiang Xu},
  year      = {2022},
  booktitle = {{The 2022 Far-field Speaker Verification Challenge (FFSVC2022)}},
  pages     = {15--19},
  doi       = {10.21437/FFSVC.2022-4},
}

@inproceedings{intra-class_min,
  author={Nam Le and Jean-Marc Odobez},
  title={{Robust and Discriminative Speaker Embedding via Intra-Class Distance Variance Regularization}},
  year=2018,
  booktitle={Proc. Interspeech 2018},
  pages={2257--2261},
  doi={10.21437/Interspeech.2018-1685}
}

@inproceedings{polyak21_interspeech,
  author={Adam Polyak and Yossi Adi and Jade Copet and Eugene Kharitonov and Kushal Lakhotia and Wei-Ning Hsu and Abdelrahman Mohamed and Emmanuel Dupoux},
  title={{Speech Resynthesis from Discrete Disentangled Self-Supervised Representations}},
  year=2021,
  booktitle={Proc. Interspeech 2021},
  pages={3615--3619},
  doi={10.21437/Interspeech.2021-475}
}

@inproceedings{hifigan,
 author = {Kong, Jungil and Kim, Jaehyeon and Bae, Jaekyoung},
 booktitle = {Advances in Neural Information Processing Systems},
 editor = {H. Larochelle and M. Ranzato and R. Hadsell and M.F. Balcan and H. Lin},
 pages = {17022--17033},
 publisher = {Curran Associates, Inc.},
 title = {HiFi-GAN: Generative Adversarial Networks for Efficient and High Fidelity Speech Synthesis},
 url = {https://proceedings.neurips.cc/paper_files/paper/2020/file/c5d736809766d46260d816d8dbc9eb44-Paper.pdf},
 volume = {33},
 year = {2020}
}

@article{hubert,
author = {Hsu, Wei-Ning and Bolte, Benjamin and Tsai, Yao-Hung Hubert and Lakhotia, Kushal and Salakhutdinov, Ruslan and Mohamed, Abdelrahman},
title = {{HuBERT}: Self-Supervised Speech Representation Learning by Masked Prediction of Hidden Units},
year = {2021},
issue_date = {2021},
publisher = {IEEE Press},
volume = {29},
issn = {2329-9290},
url = {https://doi.org/10.1109/TASLP.2021.3122291},
doi = {10.1109/TASLP.2021.3122291},
journal = {IEEE/ACM Trans. Audio, Speech and Lang. Proc.},
month = {oct},
pages = {3451–3460},
numpages = {10}
}

@article{vctk,
  author = {Yamagishi, Junichi and Veaux, Christophe and MacDonald, Kirsten},
  title = {{CSTR VCTK Corpus}: English Multi-speaker Corpus for {CSTR} Voice Cloning Toolkit (version 0.92)},
  journal = {University of Edinburgh. The Centre for Speech Technology Research (CSTR)},
  year = {2019},
  note = {doi: 10.7488/ds/2645}
}

@article{dio,
author = {Morise, Masanori and Yokomori, Fumiya and Ozawa, Kenji},
year = {2016},
month = {07},
pages = {1877-1884},
title = {WORLD: A Vocoder-Based High-Quality Speech Synthesis System for Real-Time Applications},
volume = {E99.D},
journal = {IEICE Transactions on Information and Systems},
doi = {10.1587/transinf.2015EDP7457}
}

@misc{dhariwal2020jukebox,
      title={Jukebox: A Generative Model for Music}, 
      author={Prafulla Dhariwal and Heewoo Jun and Christine Payne and Jong Wook Kim and Alec Radford and Ilya Sutskever},
      year={2020},
      eprint={2005.00341},
      archivePrefix={arXiv},
      primaryClass={eess.AS},
      url={https://arxiv.org/abs/2005.00341},
}

@INPROCEEDINGS{librispeech,
  author={Panayotov, Vassil and Chen, Guoguo and Povey, Daniel and Khudanpur, Sanjeev},
  booktitle={2015 IEEE International Conference on Acoustics, Speech and Signal Processing (ICASSP)}, 
  title={Librispeech: An ASR corpus based on public domain audio books}, 
  year={2015},
  volume={},
  number={},
  pages={5206-5210},
  doi={10.1109/ICASSP.2015.7178964}}

@inproceedings{voxceleb2,
  title     = {{VoxCeleb2: Deep Speaker Recognition}},
  author    = {Joon Son Chung and Arsha Nagrani and Andrew Zisserman},
  year      = {2018},
  booktitle = {{Interspeech 2018}},
  pages     = {1086--1090},
  doi       = {10.21437/Interspeech.2018-1929},
  issn      = {2958-1796},
}

@article{sismanoverview,
author = {Sisman, Berrak and Yamagishi, Junichi and King, Simon and Li, Haizhou},
title = {An Overview of Voice Conversion and Its Challenges: From Statistical Modeling to Deep Learning},
year = {2020},
issue_date = {2021},
publisher = {IEEE Press},
volume = {29},
issn = {2329-9290},
url = {https://doi.org/10.1109/TASLP.2020.3038524},
doi = {10.1109/TASLP.2020.3038524},
journal = {IEEE/ACM Trans. Audio, Speech and Lang. Proc.},
month = {nov},
pages = {132–157},
numpages = {26}
}

@inproceedings{berisha,
  author       = {Jianwei Zhang and
                  Suren Jayasuriya and
                  Visar Berisha},
  editor       = {Alice Oh and
                  Tristan Naumann and
                  Amir Globerson and
                  Kate Saenko and
                  Moritz Hardt and
                  Sergey Levine},
  title        = {Learning Repeatable Speech Embeddings Using An Intra-class Correlation
                  Regularizer},
  booktitle    = {Advances in Neural Information Processing Systems 36: Annual Conference
                  on Neural Information Processing Systems 2023, NeurIPS 2023, New Orleans,
                  LA, USA, December 10 - 16, 2023},
  year         = {2023},
  url          = {http://papers.nips.cc/paper\_files/paper/2023/hash/f0aa7e9e67515fa0c607c2959ccda6a0-Abstract-Conference.html},
  bibsource    = {dblp computer science bibliography, https://dblp.org}
}

@inproceedings{metric-sv,
  author={Joon Son Chung and Jaesung Huh and Seongkyu Mun and Minjae Lee and Hee-Soo Heo and Soyeon Choe and Chiheon Ham and Sunghwan Jung and Bong-Jin Lee and Icksang Han},
  title={{In Defence of Metric Learning for Speaker Recognition}},
  year=2020,
  booktitle={Proc. Interspeech 2020},
  pages={2977--2981},
  doi={10.21437/Interspeech.2020-1064}
}

@misc{speechbrain,
  title={{SpeechBrain}: A General-Purpose Speech Toolkit},
  author={Mirco Ravanelli and Titouan Parcollet and Peter Plantinga and Aku Rouhe and Samuele Cornell and Loren Lugosch and Cem Subakan and Nauman Dawalatabad and Abdelwahab Heba and Jianyuan Zhong and Ju-Chieh Chou and Sung-Lin Yeh and Szu-Wei Fu and Chien-Feng Liao and Elena Rastorgueva and François Grondin and William Aris and Hwidong Na and Yan Gao and Renato De Mori and Yoshua Bengio},
  year={2021},
  eprint={2106.04624},
  archivePrefix={arXiv},
  primaryClass={eess.AS},
  note={arXiv:2106.04624}
}

@inproceedings{wav2vec2,
author = {Baevski, Alexei and Zhou, Henry and Mohamed, Abdelrahman and Auli, Michael},
title = {wav2vec 2.0: a framework for self-supervised learning of speech representations},
year = {2020},
isbn = {9781713829546},
publisher = {Curran Associates Inc.},
address = {Red Hook, NY, USA},
booktitle = {Proceedings of the 34th International Conference on Neural Information Processing Systems},
articleno = {1044},
numpages = {12},
location = {Vancouver, BC, Canada},
series = {NIPS '20}
}

@INPROCEEDINGS{free-vc,
  author={Li, Jingyi and Tu, Weiping and Xiao, Li},
  booktitle={ICASSP 2023 - 2023 IEEE International Conference on Acoustics, Speech and Signal Processing (ICASSP)}, 
  title={Freevc: Towards High-Quality Text-Free One-Shot Voice Conversion}, 
  year={2023},
  volume={},
  number={},
  pages={1-5},
  doi={10.1109/ICASSP49357.2023.10095191}}

@inproceedings{
popov2022diffusionbased,
title={Diffusion-Based Voice Conversion with Fast Maximum Likelihood Sampling Scheme},
author={Vadim Popov and Ivan Vovk and Vladimir Gogoryan and Tasnima Sadekova and Mikhail Sergeevich Kudinov and Jiansheng Wei},
booktitle={International Conference on Learning Representations},
year={2022},
url={https://openreview.net/forum?id=8c50f-DoWAu}
}

@article{diagnostics,
author = {Cerňak, Miloš and Rusko, Milan and Trnka, Marián},
year = {2009},
month = {01},
pages = {5-9},
title = {Diagnostic evaluation of synthetic speech using speech recognition},
volume = {8},
journal = {16th International Congress on Sound and Vibration 2009, ICSV 2009}
}

@article{yamagishi_overview,
  title={A review on subjective and objective evaluation of synthetic speech},
  author={Erica Cooper and Wen-Chin Huang and Yu Tsao and Hsin-Min Wang and Tomoki Toda and Junichi Yamagishi},
  journal={Acoustical Science and Technology},
  volume={45},
  number={4},
  pages={161-183},
  year={2024},
  doi={10.1250/ast.e24.12}
}

@ARTICLE{generalized,
  author={Han, Min Hyun and Mun, Sung Hwan and Kim, Nam Soo},
  journal={IEEE Access}, 
  title={Generalized Score Comparison-Based Learning Objective for Deep Speaker Embedding}, 
  year={2025},
  volume={13},
  number={},
  pages={51194-51207},
  doi={10.1109/ACCESS.2025.3552790}}

@ARTICLE{positive_samp,
  author={Lepage, Theo and Dehak, Reda},
  journal={IEEE Transactions on Audio, Speech and Language Processing}, 
  title={Self-Supervised Frameworks for Speaker Verification via Bootstrapped Positive Sampling}, 
  year={2025},
  volume={33},
  number={},
  pages={2932-2945},
  doi={10.1109/TASLPRO.2025.3587462}}

@ARTICLE{max_sep,
  author={Li, Zhe and Mak, Man-Wai and Pilanci, Mert and Meng, Helen},
  journal={IEEE Transactions on Audio, Speech and Language Processing}, 
  title={Mutual Information-Enhanced Contrastive Learning With Margin for Maximal Speaker Separability}, 
  year={2025},
  volume={33},
  number={},
  pages={2961-2972},
  doi={10.1109/TASLPRO.2025.3583485}}

@article{oytun_overview,
  title = {Deep learning-based expressive speech synthesis: a systematic review of approaches, challenges, and resources},
  author = {Huda Barakat and Oytun T{\"u}rk and Cenk Demiroğlu},
  journal = {EURASIP J. Audio Speech Music. Process.},
  year = {2024},
  volume = {2024},
  pages = {11}
}

\endgroup
\end{document}